\documentclass[b5paper,twoside,reqno]{bjp}
\usepackage{graphics}
\usepackage[pdftex]{graphicx}
\usepackage{amssymb,amsmath}
\usepackage{times}
\renewcommand\congress{\small\sf Section: Theoretical Physics}

\newcommand\be{\begin{equation}}
\newcommand\ee{\end{equation}}
\newcommand{\bea}{\begin{eqnarray}}
\newcommand{\eea}{\end{eqnarray}}

\newcommand{\nn}{\nonumber}
\newcommand{\pd}{\partial}

\begin{document}

\sloppy \raggedbottom

 \setcounter{page}{1}



\title{On Slow-roll Glueball Inflation from Holography}

\runningheads{Slow-roll Glueball Inflation from Holography}{L.~Anguelova}

\begin{start}
\author{L.~Anguelova}{}

\address{Institute of Nuclear Research and Nuclear Energy,
BAS, Sofia 1784, Bulgaria}{}


\begin{Abstract}
We investigate glueball inflation model-building via the methods of the gauge/gravity duality. For that purpose, we consider a certain 5d consistent truncation of type IIB supergravity. This theory admits a solution, whose metric is of the form of a $dS_4$ fibration over a fifth dimension. We find a new time-dependent deformation around this solution, which allows for a small $\eta$ parameter of the corresponding inflationary model. This resolves a problem with a previous solution that allowed only $\eta$ of order one and thus gave only an ultra-slow roll regime, but not regular slow roll.
\end{Abstract}

\end{start}

\section[]{Introduction}

Modern developments, originating from string theory, have provided a powerful tool for studying the nonperturbative regime of certain gauge theories. The idea is that at strong coupling those gauge theories are dual to gravitational backgrounds in a different number of dimensions. The first example of such a duality is the AdS/CFT correspondence proposed in \cite{JM}. Since then, many more dualities have been explored and this broad area of research is now known as the Gauge/Gravity Duality. This has enabled important advances in the study of the quark-gluon plasma, high temperature superconductivity and dynamical chiral symmetry breaking, for example. Our goal is to apply this powerful method to the investigation of composite Cosmological Inflation.

Inflationary models, in which the inflaton is a composite state in a strongly-coupled gauge sector \cite{CJS,BCJS} instead of a fundamental scalar, are of interest because they are generically expected to provide a resolution to the long-standing inflationary $\eta$ problem \cite{CLLSW,DRT}. We will aim to build gravitational duals of such models by looking for a certain kind of solutions to the five-dimensional consistent truncation of type IIB established in \cite{BHM}. This effective 5d theory is a very useful unifying framework in Gauge/Gravity Duality, since it encompasses many famous gravity duals, like the Maldacena-Nunez \cite{MN} and Klebanov-Strassler \cite{KS} solutions dual to ${\cal N} = 1$ SYM, together with their deformations \cite{NPP,ENP,EGNP} that correspond to multi-scale dynamics on the gauge theory side.

In \cite{ASW, ASW2} we found non-supersymmetric solutions of this 5d theory, which are $dS_4$ fibrations over a fifth (radial) dimension. And in \cite{LA, LA2}, we considered a certain small deformation around the $dS_4$ component of these solutions, such that the 4d nearly de Sitter space-time has a time-dependent Hubble parameter. Such 5d backgrounds would represent gravitational duals to glueball inflation models, since their 5d metric is supported by nontrivial 5d scalars that correspond to glueballs in the dual strongly-coupled gauge theory. Unfortunately, however, the solution found in \cite{LA} was too restrictive and, as a consequence, corresponding only to an ultra-slow roll regime. To obtain a full inflationary model, one would like to have a dual of regular slow roll. We suggested in \cite{LA2} that the unwanted restriction might be removed, thus opening the possibility for obtaining duals of slow roll inflationary models, by considering the next order in an expansion in a certain small parameter. Here we will show that this is indeed the case.

\section[]{The gauge/gravity duality construction}

The action of interest for us is the following:
\be \label{S5d}
S = \int d^5 x \sqrt{- det g} \left[ - \frac{R}{4} + \frac{1}{2} \,G_{ij} (\Phi) \,\pd_{I} \Phi^i \pd^{I} \Phi^i + V (\Phi) \right] \, ,
\ee
where $\{ \Phi^i \}$ is a set of five-dimensional scalars, $g_{IJ}$ is the 5d spacetime metric and $R$ is its Ricci scalar. This action arises from a particular consistent truncation of ten-dimensional type IIB supergravity established in \cite{BHM}. For more details on the scalar potential $V(\Phi)$ and the sigma-model metric $G_{ij} (\Phi)$, see \cite{BHM} or the concise summary in \cite{ASW}. The subtruncation studied in \cite{ASW} has zero NS flux and thus six five-dimensional scalars:
\be \label{6scalars}
\{\Phi^i (x^I)\} = \{ p (x^I), x (x^I), g (x^I), \phi (x^I), a(x^I), b(x^I) \} \,\, ,
\ee
where we have used the same notation as in \cite{ASW}. The equations of motion, that follow from (\ref{S5d}), are:
\bea \label{EoM}
\nabla^2 \Phi^i + {\cal G}^i{}_{jk} \,g^{IJ} (\pd_I \Phi^j) (\pd_J \Phi^k) - V^i &=& 0 \quad , \nn \\
- R_{IJ} + 2 \,G_{ij} \,(\pd_I \Phi^i) (\pd_J \Phi^j) + \frac{4}{3} \,g_{IJ} V &=& 0 \quad ,
\eea
where $\nabla^2 = \nabla_I \nabla^I$ and $V^i = G^{ij} V_j$ with $V_j \equiv \frac{\pd V}{\pd \Phi^j}$\,, while ${\cal G}^i{}_{jk}$ are the Christoffel symbols for the metric $G_{ij}$\,.

A significant number of important gravity duals of strongly-coupled supersymmetric gauge theories, living in 4d Minkowski space, can be found as solutions of the system (\ref{EoM}) with a 5d metric of the form:
\be \label{5dmetr}
ds_5^2 = e^{2 A(z)} \eta_{\mu \nu} dx^{\mu} dx^{\nu} + dz^2 \,\, ,
\ee
where $\mu, \nu = 0,1,2,3$\,, \,and with the fields $\Phi^i$ being functions only of the radial variable $z$, i.e. $\Phi^i = \Phi^i (z)$. This includes the famous Maldacena-Nunez \cite{MN} and Klebanov-Strassler \cite{KS} solutions, as well as a variety of recently found deformations of theirs \cite{NPP,ENP,EGNP}. The distinction between different solutions is in the form of the  functions $A(z)$ and $\Phi^i (z)$ that characterize them. In this gauge/gravity duality context, the 5d scalars $\Phi^i$ describe glueballs in the dual 4d gauge theory. The Maldacena-Nunez solution, for example, can be obtained for \cite{BHM,PT}:
\be
Q=0 \quad , \quad b=a \quad , \quad x=\frac{1}{2} g - 3p \quad , \quad \phi = -6p - g - 2 \ln P \,\, ,
\ee
and certain expressions for the functions $a(z)$, $g(z)$ and $p(z)$; here $P$ and $Q$ are the constant overall coefficients in the 10d RR forms $F_3$ and $F_5$ respectively.

One can search for gravity duals of strongly-coupled gauge theories living in a curved 4d spacetime by modifying the metric ansatz (\ref{5dmetr}) to:
\be \label{5dmetrCurved}
ds_5^2 = e^{2 A(z)} g_{\mu \nu} dx^{\mu} dx^{\nu} + dz^2 \,\, ,
\ee
where $g_{\mu \nu}$ is a four-dimensional metric. In \cite{ASW}, we analyzed the cases of $g_{\mu \nu}$ being de Sitter or anti-de Sitter. Working within a consistent subtruncation such that three of the six scalars in (\ref{6scalars}) are identically vanishing, namely:
\be
g (x^I) \equiv 0 \quad , \quad a (x^I) \equiv 0 \quad , \quad b (x^I) \equiv 0 \quad ,
\ee
we were able to find three families of solutions to the system (\ref{EoM}) with a 4d de Sitter metric $g_{\mu \nu}$. More precisely, in those solutions the 5d metric has the form \cite{ASW}:
\be \label{metricans}
ds_5^2 = e^{2A(z)} \left[ -dt^2 + e^{2{\cal H}t} d\vec{x}^2 \right] + dz^2 \,\, ,
\ee
where ${\cal H} = const$ is the 4d Hubble constant. One of these families of solutions was found analytically, while the other two were obtained only numerically. 

To build a gravity dual of a model of glueball inflation, we would like to have solutions of (\ref{EoM}) with a 5d metric of the form (\ref{metricans}), {\it but} with a time-dependent Hubble parameter ${\cal H} = {\cal H} (t)$. Since one of the slow-roll conditions of inflationary cosmology is (see \cite{DB}, for example): 
\be
- \frac{\dot{{\cal H}}}{{\cal H}^2} <\!\!< 1 \, ,
\ee
the desired gravity duals can be viewed as small time-dependent deformations around the solutions of \cite{ASW}, which have ${\cal H} = const$. For technical convenience, we will only look for deformations around the analytical solution. The latter has the form \cite{ASW}:
\bea \label{ZeroOrSol}
A_0 (z) &=& \ln (z+C) + \frac{1}{2} \,\ln \!\left( \frac{7}{3} {\cal H}_0^2 \right) \,\, , \nn \\
p_0 (z) &=& - \frac{1}{7} \ln (z+C) - \frac{1}{14} \,\ln \!\left( \frac{7 N^2}{9} \right) \,\, , \nn \\
x_0 (z) &=& - 6 \,p_0 (z) \qquad , \qquad \phi_0 = 0 \quad ,
\eea
where $C$ and $N$ are constants and we have used the subscript $0$ to denote that these functions are specific to the analytical solution.

Now, we would like to look for solutions of the system (\ref{EoM}), which are small deformations around (\ref{ZeroOrSol}). Note that a generic deformation of (\ref{metricans}), preserving the same basic structure, has the form:
\be \label{TimeDepMetricAnz}
ds_5^2 = e^{2A(t,z)} \left[ -dt^2 + e^{2H(t,z)} d\vec{x}^2 \right] + dz^2 \,\, . 
\ee
To search for small deformations, let us introduce a parameter $\gamma$ satisfying
\be
\gamma <\!\!< 1 \,\, ,
\ee
and consider expansions in powers of $\gamma$, around the zeroth order solution (\ref{ZeroOrSol}), for the metric functions in (\ref{TimeDepMetricAnz}) and the relevant 5d scalars $\Phi^i$.  
Namely, let us make the following ansatz for the nonvanishing 5d fields:
\be \label{p0x0}
p (t,z) = p_0 (z) \quad , \quad x (t,z) = x_0 (z) \,\,\, ,
\ee
and
\bea \label{Expansions}
\phi(t,z) &=& \gamma \,\phi_{(1)} (t,z) + \gamma^3 \phi_{(3)} (t,z) + {\cal O} (\gamma^5) \,\,\, , \nn \\
A(t,z) &=& A_0(z) + \gamma^2 A_{(2)}(t,z) + \gamma^4 A_{(4)} (t,z) + {\cal O} (\gamma^6) \,\,\, , \nn \\
H(t,z) &=& {\cal H}_0 \,t + \gamma^2 H_{(2)} (t,z) + \gamma^4 H_{(4)} (t,z) + {\cal O} (\gamma^6) \,\,\, .
\eea
Clearly, (\ref{p0x0}) means that the scalar fields $p$ and $x$ are kept exactly the same as in the zeroth order solution (\ref{ZeroOrSol}). Thus, the only deformed scalar (and, therefore, potential inflaton) is $\phi$. This choice of ansatz is not necessary. However, it is technically very convenient, because the zeroth order solution $\phi_0$ vanishes (see (\ref{ZeroOrSol})) and also because $\phi$ is a flat direction of the potential $V(\Phi)$ (see \cite{LA}) and thus $V^{\phi} = 0$. Note also that expanding the metric functions $A$ and $H$ in different powers of $\gamma$, compared to $\phi$, will play a key role in enabling us to solve analytically the highly involved coupled system of field equations.

\section[]{A new solution at order $\gamma^4$}

In \cite{LA}, using the ansatz (\ref{TimeDepMetricAnz})-(\ref{Expansions}), we solved the system (\ref{EoM}) to leading order in $\gamma$, i.e. to ${\cal O}(\gamma)$ for $\phi$ and to ${\cal O}(\gamma^2)$ for $A$ and $H$. The result turned out to be quite restrictive and, as a consequence, providing a gravity dual of only a model of so called ultra-slow roll inflation. This is a short-lived regime of limited cosmological interest, as it cannot by itself produce enough inflationary expansion. To find a solution, corresponding to regular slow-roll, here we will explore the equations of motion at the next order in $\gamma$ in the ansatz (\ref{Expansions}), namely at ${\cal O} (\gamma^3)$ for the scalar (inflaton) $\phi$ and at ${\cal O}(\gamma^4)$ for the metric warp factors $A$ and $H$. 

Let us begin with the following observation. One can immediately see by inspection of (3.14) in \cite{LA} that the equations of motion to leading order in $\gamma$ are also solved if one takes:
\be \label{phi1A2H2}
\phi_{(1)} = 0 \qquad , \qquad A_{(2)} = 0 \qquad , \qquad H_{(2)} = 0 \,\,\, .
\ee
In that case, the leading deformations arise at the next order:
\bea \label{Expansions2}
\phi(t,z) &=& \gamma^3 \,\phi_{(3)} (t,z) + {\cal O} (\gamma^5) \,\,\, , \nn \\
A(t,z) &=& A_{(0)}(z) + \gamma^4 A_{(4)}(t,z) + {\cal O} (\gamma^6) \,\,\, , \nn \\
H(t,z) &=& H_{(0)} (t) + \gamma^4 H_{(4)} (t,z) +{\cal O} (\gamma^6) \,\,\, ,
\eea
where $H_{(0)} (t) = {\cal H}_0 \,t$. For future convenience, we will denote from now on:
\be
{\cal H}_0 \equiv h \,\,\, .
\ee
Working with (\ref{phi1A2H2})-(\ref{Expansions2}) is very useful, because now $\phi^2 \sim {\cal O} (\gamma^6)$ or higher and thus the scalar $\phi(t,z)$ does not enter the ${\cal O} (\gamma^4)$ equations for $A_{(4)}$ and $H_{(4)}$. This decoupling allows significantly more freedom in inflation model building. Indeed, in \cite{LA} we ended up with the restrictive ultra-slow roll regime, precisely because there $\phi \sim {\cal O}(\gamma^2)$ and thus it entered the ${\cal O}(\gamma^2)$ equations for $A_{(2)}$ and $H_{(2)}$. This mixing between the inflaton $\phi$ and the functions $A$, $H$ led to the restrictions that ruled out a regular slow roll solution in that case. So, since we would like to relax those restrictions, now we will explore the consequences of the ansatz (\ref{phi1A2H2})-(\ref{Expansions2}).

Substituting (\ref{Expansions2}) in (\ref{EoM}), it is easy to find that the order $\gamma^4$ metric equations are:
\bea \label{E1E2E3E4}
&&E1: \quad - h^2 \!\left( \frac{7}{3} (z+C)^2 A_{(4)}'' + \frac{56}{3} (z+C) A_{(4)}' + 7 (z+C) H_{(4)}' + 6 A_{(4)} \right) \nn \\
&&\hspace*{1.3cm}+ h \left( 3 \dot{A}_{(4)} + 6 \dot{H}_{(4)} \right) + 3 \ddot{A}_{(4)} + 3 \ddot{H}_{(4)} = 0 \,\, , \nn \\
&&E2: \quad h^2 \!\left( \frac{7}{3} (z+C)^2 \left[ A_{(4)}'' + H_{(4)}'' \right] + \frac{56}{3} (z+C) A_{(4)}' + \frac{49}{3} (z+C) H_{(4)}' \right. \nn \\
&&\hspace*{1.3cm}+ \,6 A_{(4)} \bigg) - h \left( 5 \dot{A}_{(4)} + 6 \dot{H}_{(4)} \right) - \ddot{A}_{(4)} - \ddot{H}_{(4)} = 0 \,\, , \nn \\ 
&&E3: \quad 4 A_{(4)}'' + 3 H_{(4)}'' + \frac{2}{z+C} \left( 4 A_{(4)}' + 3 H_{(4)}' \right) = 0 \,\, , \nn \\
&&E4: \quad 3 \dot{A}_{(4)}' + 3 \dot{H}_{(4)}' + 3 \,h H_{(4)}' = 0 \,\, ,
\eea
where \,$\dot{} \equiv \partial_t$ and $' \equiv \partial_z$. This is exactly the same as (3.14) of \cite{LA} but with the $\phi$ terms dropped and the index $(2)$ of the functions $A$, $H$ substituted by $(4)$. Inspired by the way the system was solved in \cite{LA} at the lower order in $\gamma$, let us now look for solutions such that:
\be
A_{(4)}' = 0 \qquad {\rm and} \qquad H_{(4)}' = 0 \,\, .
\ee
In this case, equations $E3$ and $E4$ are identically satisfied. So we are left with the following two equations:
\bea \label{E1E2}
&&E1: \qquad - 6 \,h^2 A_{(4)} + h \left( 3 \dot{A}_{(4)} + 6 \dot{H}_{(4)} \right) + 3 \ddot{A}_{(4)} + 3 \ddot{H}_{(4)} = 0 \nn \\
&&E2: \qquad 6 \,h^2 A_{(4)} - h \left( 5 \dot{A}_{(4)} + 6 \dot{H}_{(4)} \right) - \ddot{A}_{(4)} - \ddot{H}_{(4)} = 0 \,\, .
\eea
Then, considering $E1+3E2$, we find:
\be \label{H4eq}
\dot{H}_{(4)} = h A_{(4)} - \dot{A}_{(4)} \,\, .
\ee 
It is easy to verify, by substituting (\ref{H4eq}) back into (\ref{E1E2}), that both equations there are identically satisfied. So (\ref{H4eq}) is the only metric equation left to solve.

An easy way of solving this master equation is to take:
\be \label{Asol}
A_{(4)} = C_a e^{\omega t} \,\, ,
\ee
where $C_a,\omega = const$. Substituting this in (\ref{H4eq}), we find:
\be \label{Hsol}
H_{(4)} = \frac{h-\omega}{\omega} \,C_a \,e^{\omega t} + const \,\, .
\ee
This solution is in the spirit of the one for $A_{(2)}$ and $H_{(2)}$ found in \cite{LA}. Note, however, the crucial difference that now the constant $\omega$ is completely independent of, and unrelated to, any constants that we will encounter below in solving for the inflaton $\phi$, unlike the situation in \cite{LA}.

Now let us consider the $\phi_{(3)}$ equation of motion, which arises at order $\gamma^3$ and is completely decoupled from the metric deformations, just as was the case for $\phi_{(1)}$ in \cite{LA}. Upon substituting (\ref{phi1A2H2}) in (A.2) of \cite{LA}, one can immediately see that the $\phi$ field equation is:
\be \label{phi3}
\ddot{\phi}_{(3)} + 3 \,h \,\dot{\phi}_{(3)} = e^{2 A_{(0)}} \!\left( \phi''_{(3)} + 4 A'_{(0)} \phi'_{(3)} \right) \,\, ,
\ee
which is exactly the same as (3.2) of \cite{LA} but with the index $(1)$ substituted by $(3)$. Hence, we can find a solution to (\ref{phi3}) in exactly the same way as for equation (3.2) of \cite{LA}. Namely, making the ansatz
\be \label{phifac}
\phi_{(3)} = \Phi_1 (t) \,\Phi_2 (z)
\ee
and considering the eigen problems
\be \label{EigenTZ}
\ddot{\phi}_{(3)} + 3 h \dot{\phi}_{(3)} = \lambda \,\phi_{(3)} \quad {\rm and} \quad e^{2 A_{(0)}} \!\left( \!\phi_{(3)}'' + 4 A'_{(0)} \phi_{(3)}' \!\right) = \lambda \,\phi_{(3)}
\ee
with $\lambda = const$, we obtain: 
\be \label{Phi1tsol}
\Phi_1 (t) = C_1 \,e^{k_+ t} + C_2 \,e^{k_- t} \qquad {\rm with} \qquad k_{\pm} = -\frac{3 h}{2} \pm \frac{\sqrt{9 h^2 + 4 \lambda} }{2}
\ee
and 
\be \label{PhizC3C4Sol}
\Phi_2 (z) = C_3 (z+C)^{\alpha_+} + C_4 (z+C)^{\alpha_-} \quad {\rm with} \quad \alpha_{\pm} = - \frac{3}{2} \pm \frac{3}{2} \sqrt{1 + \frac{4 \,\lambda}{21 \,h^2}} \,\, .
\ee
Note that, in principle, we should be able to compute the dynamically fixed mass of the inflaton  (or rather, find a discrete spectrum for it) by studying fluctuations around the $\gamma$-deformed background solution obtained here and imposing suitable boundary conditions. This is in the same vein as computing the glueball spectrum in, for example, the work \cite{ASW3}. However, the present situation is significantly more involved as the fluctuation equations, used in \cite{ASW3}, were derived in \cite{DE} only for fluctuations around a time-independent background. Here though we are interested in inflationary backgrounds, which are time-dependent. So the first step, toward finding the dynamically fixed mass of the inflaton, is to derive the appropriate fluctuation equations around a time-dependent background. This is a rather complicated task, which begins with formulating suitable gauge-invariant fluctuation variables that mix the fluctuations of the various scalar degrees of freedom involved.\footnote{In that regard, note that the superficial similarity of $\lambda$ to a mass parameter, because of the way it enters (\ref{EigenTZ}), is deceptive since the equations that the fluctuation fields satisfy will certainly be different. In particular, the fluctuation for $\phi$ is not a gauge-invariant quantity by itself, as the $\phi$-background solution is not constant (for a discussion of that issue see \cite{ASW3}). Therefore, the appropriate gauge-invariant variable will not be just a fluctuation of $\phi$ around the $\gamma$-deformed $\phi$-background found here.} We hope to address this problem in more detail in the near future.

\section[]{Discussion}

Here we explored the idea of looking for more general solutions then the one in \cite{LA} by going to a higher order in the small parameter $\gamma$ in the expansions (\ref{Expansions}). Making the ansatz (\ref{phi1A2H2}), so that the leading deformations of the metric functions are at order $\gamma^4$, as is obvious in (\ref{Expansions2}), we derived a single master equation (\ref{H4eq}), whose solutions are guaranteed to satisfy the whole coupled system of metric equations of motion. This equation encodes a particular class of promising solutions. An example, bearing similarity to the metric functions in \cite{LA}, is given by (\ref{Asol})-(\ref{Hsol}). It is interesting to look for other solutions of (\ref{H4eq}), as well as for more general solutions of the full system. In any case however, the great advantage of the considerations here is that the metric field equations are entirely decoupled from the inflaton. 

The inflaton equation can be solved similarly to \cite{LA} with the result being (\ref{phifac}) and (\ref{Phi1tsol})-(\ref{PhizC3C4Sol}). The important difference with \cite{LA} is that now the constants $k_{\pm}$ in the time-dependent factor of $\phi$ are not constrained by any relation to constants appearing in the solutions for the metric functions. So, as long as $\lambda$ is arbitrary, they are arbitrary as well. This is of great significance for relaxing the problematic constraint in \cite{LA} that the second slow roll parameter $\eta$ had to be of order 1 (more precisely, $\eta = 3$). This constraint implied that the solution found there had to give ultra-slow roll, as opposed to regular slow roll inflation for which $\eta <\!\!< 1$. Let us see how this can be avoided in the present case. First, recall that by definition \cite{DB}:
\be \label{etapar}
\eta = - \frac{\ddot{\phi}}{{\cal H} \dot{\phi}} \,\,\, ,
\ee
where ${\cal H}$ and $\phi$ are the 4d Hubble parameter and the inflaton respectively. Clearly, the $z$-dependent factor in (\ref{phifac}) cancels out of the ratio in (\ref{etapar}). So, for brevity and convenience, from now on we will only keep track of the time dependent one. Hence, to leading order in the $\gamma$ deformation, we have for the inflaton:
\be
\phi = \gamma^3 \phi_{(3)} = \gamma^3 \left( C_1 e^{k_+ t} + C_2 e^{k_- t} \right) \,\, ,
\ee
where $k_{\pm}$ is given in (\ref{Phi1tsol}). Therefore, regardless of the form of the leading ${\cal O}(\gamma^4)$ correction to the metric functions, we have that the leading order result for $\eta$ is:
\be
\eta = - \, \frac{1}{h} \, \frac{C_1 k_+^2 e^{k_+ t} + C_2 k_-^2 e^{k_- t}}{C_1 k_+ e^{k_+ t} + C_2 k_- e^{k_- t}} + {\cal O} (\gamma^4) \,\, .
\ee
It is easy to see that this result allows us to obtain small $\eta$ for suitably chosen parameters. For example, let us take the integration constant $C_2 = 0$. Then, to leading order in $\gamma$, we have:
\be
\eta = - \frac{k_+}{h} = \frac{3}{2} \left( 1 - \sqrt{1 + \frac{4 \lambda}{9 h^2}} \right) \,\, .
\ee
Since $\lambda$ is a free parameter, we can choose it so that the slow roll condition $\eta <\!\!< 1$ is satisfied.

\section*{Acknowledgments}
I would like to thank P. Suranyi and L.C.R. Wijewardhana for useful discussions. I am also grateful for partial support from the European COST Action MP-1210 and the Bulgarian NSF grant DFNI T02/6 during the completion of this work.

\end{document}